\documentstyle[prl,aps,epsf]{revtex}

\begin{document}

\advance\textheight by 0.2in%%%

\draft
\twocolumn[\hsize\textwidth\columnwidth\hsize\csname@twocolumnfalse%%%%
\endcsname  %%%% these lines give two column format

\title{Depinning Transition of a Two Dimensional Vortex Lattice in a
Commensurate Periodic Potential}

\author{Violeta Gotcheva and S. Teitel}

\address{Department of Physics and Astronomy, University of Rochester,
Rochester, NY 14627}

\date{\today}

\maketitle

\begin{abstract}

We use Monte Carlo simulations of the 2D one component Coulomb gas on
a triangular lattice, to study the depinning transition of a 2D
vortex lattice in a commensurate periodic potential.  A detailed
finite size scaling analysis indicates this transition to be first
order.  No significant changes in behavior were found as vortex
density was varied over a wide range.

\end{abstract}


]

The theory of defect mediated melting of a two dimensional (2D) solid,
introduced by Kosterlitz and Thouless (KT) \cite{R1} and developed by Nelson
and Halperin (NH) \cite{R2} and Young \cite{R3} in the late 70's, has 
remained a topic of active investigation.  Numerous numerical studies 
have reported conflicting results as to whether the 2D melting transition 
is indeed a 2nd order KT type transition, or whether it is 1st 
order \cite {R4}.  The most recent results have supported the 
KT scenario \cite{R5}.  In the same paper \cite{R2}
in which they developed this
theory of 2D melting in a continuum, Nelson and 
Halperin also considered the case of a 2D solid in a commensurate 
periodic potential.  In this case, they argued that the 2D solid 
would have two distinct phases: a ``pinned'' solid with long range 
translational correlations at low temperatures, and a ``floating'' 
solid with algebraic translational correlations, similar to that found 
in the continuum, at intermediate temperatures.  By mapping the 
problem to a vector Coulomb gas, they argued that this depinning 
transition was a 2nd order KT type transition similar to that of melting,
with a universal discontinuous jump in the exponent 
$\eta$ characterizing the algebraic correlations of the floating 
solid phase.

Despite the wide attention given to the continuum 2D melting problem,
this depinning transition has been very little studied.  Only 
recently has the floating solid phase been observed in numerical 
studies of 2D vortex lattices \cite{R6}, and related XY models 
\cite{R7}.  In this paper we present the first detailed finite size 
scaling analysis of this depinning transition.  We treat the specific 
case of logarithmically interacting 2D vortices.  This is not only
a system of considerable recent interest in connection with high 
temperature superconductors, but also has a unique advantage for 
numerical simulations: it is an incompressible system.  We can 
simulate at constant density, yet there will still be a sharp transition 
temperature, rather than the finite temperature interval of coexisting phases 
that one would have for other interactions (should the transition be
1st order).  We can therefore 
avoid the controversy, that arose in 2D melting simulations, as to whether 
one should use a constant pressure rather than a constant volume 
ensemble.  For this vortex system, our results are quantitatively 
consistent with a 1st order depinning transition.

Our model is the one component classical Coulomb gas on a triangular 
lattice, given by the Hamiltonian,

%******
\begin{equation}
    {\cal H} = {1\over 2}\sum_{i,j}(n_i-f)G_{ij}(n_j-f)\enspace .
\label{e1}
\end{equation}
%*****
%
Here $n_i=0$, $1$ is the integer charge at site $i$ of an $L\times L$ 
periodic triangular lattice; $-f$ is a uniform background charge 
density; charge neutrality fixes the number of integer charges to 
$\sum_i n_i = fL^2$.  $G_{ij}$ is the 2D Coulomb potential for a 
discrete triangular lattice with periodic boundary conditions 
\cite{R6}, and the sum is over all pairs of sites.  
$G_{ij}\sim\ln |{\bf r}_i-{\bf r}_j|$ for distances large compared to 
the lattice spacing, but small compared to $L$.  The charges $n_i$ 
model logarithmically interacting vortices in the phase of a 
superconducting wavefunction, $f$ is the number of applied magnetic 
flux quanta per unit cell of the triangular lattice, and restricting 
the vortices to the sites of the triangular lattice models the 
periodic pinning potential \cite{R6}.

We study the Hamiltonian (\ref{e1}) using Monte Carlo (MC)
simulations.  Details of our simulation methods follow those 
described in Ref.\cite{R6}.  Our main results are for the fixed vortex 
density of $f=1/100$.

\begin{figure}
\epsfxsize=3.2truein
\epsfbox{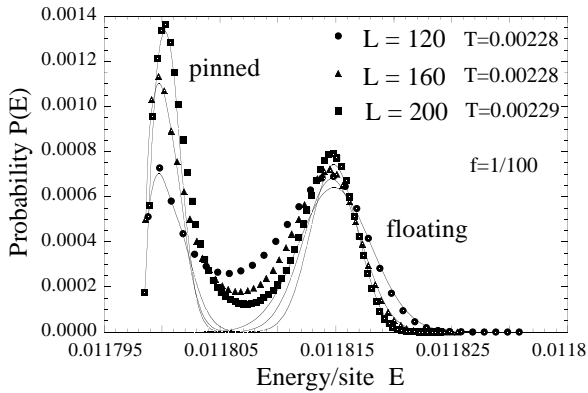}
\vspace{9pt}
\caption{
Histograms of energy density $P(E)$ vs. $E$ for vortex density 
$f=1/100$, and system sizes $L=120$, $160$ and $200$, at 
temperatures near the depinning transition $T_c$.  The two peaks 
correspond to the coexisting pinned and floating solid phases.  The peaks 
sharpen as $L$ increases, consistent with a 1st order transition.  The solid 
lines are fits to the peaks, as described in the text.
}
\label{f1}
\end{figure}

To decide the order of the depinning transition at $T_c$, we study the finite 
size scaling of the energy density histogram $P(E)$ measured close to $T_c$.  
For a first order transition, one expects to find a bimodal $P(E)$, 
with peaks that sharpen to two separated $\delta-$functions as 
$L\to\infty$; these give the differing energies of the two coexisting 
phases.  In Fig.\,\ref{f1} we plot the normalized $P(E)$ for sizes 
$L=120$, $160$, 
and $200$, near $T_c\simeq 0.00229$.  Analysis of the vortex structure 
function,
%
%*****
\begin{equation}
    S({\bf k}) = {1\over fL^2}\sum_{i,j}\langle{\rm e}^{i{\bf k}\cdot 
    ({\bf r}_i-{\bf r}_j)}\rangle\enspace,
\label{e2}
\end{equation}
%***********
%
clearly shows that the lower (upper) energy peak corresponds to the pinned
(floating) vortex lattice.  
For $L=120$, $160$, and $200$ respectively, we have carried out
$0.77$, $1.5$ and $4.7\times 10^8$ MC passes though the lattice to 
compute averages.  
These resulted in $2219$, $1186$ and $841$ hops, respectively, between 
the two peaks.
Having a large number of hops is important to achieve good 
equilibration of the relative weights of the two peaks (we have also 
done simulations for $L=80$, $100$, $140$, and $180$).
Qualitatively, we see that these peaks 
sharpen as $L$ increases, consistent with a 1st order transition.  
We now make this observation quantitative using several different 
criteria.

\begin{figure}
\epsfxsize=3.2truein
\epsfbox{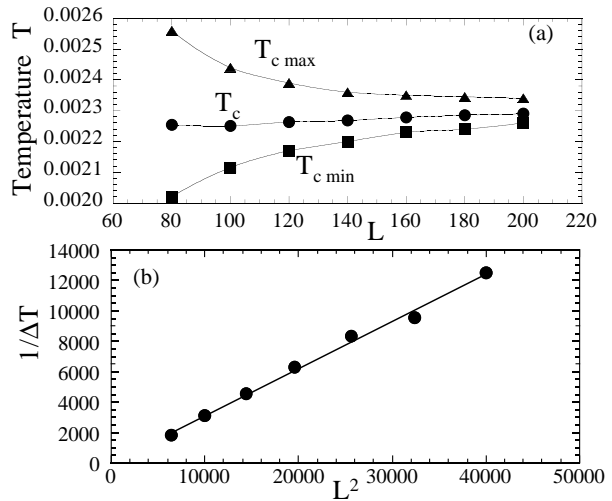}
\vspace{9pt}
\caption{
(a) Plot of the upper and lower limits, $T_{c\,{\rm max}}$ and 
$T_{c\,{\rm min}}$, of the coexistence region, and the finite size 
depinning temperature $T_c$, vs. system size $L$. (b) Inverse of the width of 
the coexistence region, $1/\Delta T=1/(T_{c\,{\rm max}}-T_{c\,{\rm min}})$, vs. 
$L^2$.  Vortex density is $f=1/100$.  The straight line is a least squares fit, and the good agreement 
is consistent with a 1st order transition.
}
\label{f2}
\end{figure}

For an infinite system, the bimodal $P(E)$ should exist only precisely at 
$T_c$, where the two phases coexist.  For finite $L$, this 
coexistence region where $P(E)$ is noticeably bimodal persists over a 
finite temperature interval $\Delta T$.  Since the relative weight of 
each peak is determined by the total free energy difference between 
the two phases, $\Delta F\sim L^d$, we expect the scaling $\Delta 
T\simeq 1/\Delta F = 1/L^d = 1/L^2$ in 2D.  Using standard 
methods \cite{R8} to extrapolate the histograms of Fig.\,\ref{f1} to 
nearby temperatures, we define the upper (lower) limit of the 
coexistence region, $T_{c\, {\rm max\,(min)}}$, as the temperature at which 
the height of the lower (upper) peak has decreased to $1\%$ the height 
of the upper (lower) peak.  In Fig.\,\ref{f2} we plot the results.  We 
see that $T_{c\, {\rm max}}$ and $T_{c\, \rm min}$ converge to a 
common value as $L$ 
increases, and that $\Delta T\sim 1/L^2$ as expected for a 1st order 
transition.

\begin{figure}
\epsfxsize=3.2truein
\epsfbox{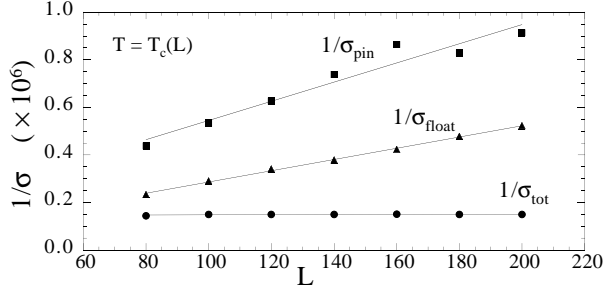}
\vspace{9pt}
\caption{
Inverse of the total width $\sigma_{\rm tot}$ of the energy 
histogram $P(E)$, and inverse of the widths, $\sigma_{\rm pin}$ and
$\sigma_{\rm float}$, of the pinned and floating solid peaks, vs. 
system size $L$.    Vortex density is $f=1/100$.
The straight lines are least squares fits.
The observed scaling $1/\sigma_{\rm tot}\sim$ constant, $1/\sigma_{\rm 
pin},1/\sigma_{\rm float}\sim L$ is consistent with a 1st order 
transition.
}
\label{f3}
\end{figure}

\begin{figure}
\epsfxsize=3.2truein
\epsfbox{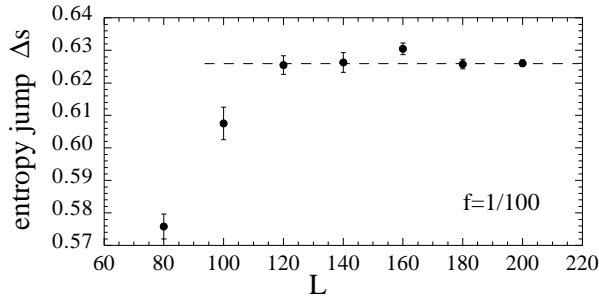}
\vspace{9pt}
\caption{
Entropy jump {\it per vortex}, $\Delta s = (\langle E\rangle_{\rm float}-\langle 
E\rangle_{\rm pin})/(fT_c)$,
at the depinning transition vs. system size $L$.
The vortex density is $f=1/100$.  The dashed line is a guide
to the eye only.
}
\label{f4}
\end{figure}

Next, we consider in detail how the two peaks sharpen as $L$ increases.
For a 1st order transition at $T_c$, the width $\sigma$ 
of each peak is determined 
by ordinary {\it non-critical} finite size fluctuations giving, 
$\sigma\sim 1/L^{d/2}=1/L$ in 2D.  The total width of the bimodal $P(E)$, 
$\sigma_{\rm tot}$, approaches a constant (this is equivalent to 
the familiar observation that the specific heat $c\equiv L^d\sigma_{\rm 
tot}/T\sim L^d$ at a 1st order transition).  To verify these scaling 
behaviors, we need to deconvolve the $P(E)$ of Fig.\,\ref{f1} into 
two separate peaks, and to determine $T_c(L)$ by the criteria 
that the two peaks subtend equal areas \cite{R9}.  For large enough 
$L$, we expect each peak to have a Gaussian shape, and we have found 
this to give a good fit for the floating solid peak.  The pinned 
solid peak however lies too close to the ground state energy $E_0$; even for 
our biggest size $L=200$, the states in this peak correspond to a few 
discrete excitations above the ground state.  We therefore fit the 
pinned solid peak to the {\it ad hoc} form, 
$(E-E_0){\rm exp}[-(E-E_1)^2/2\sigma^2]$, where $E_1$ and $\sigma$ are 
fitting parameters. Using data from only the far side of each peak, we 
fit $P(E)$ to the sum of a Gaussian and the modified Gaussian above, 
to get the fitted curves shown in Fig.\,\ref{f1}.  From these fits 
we determine the probabilities that a state with a given value 
of $E$ belongs to the pinned phase, the floating phase, or neither (the 
later being the ``transition'' states, consisting of large domains 
of one phase in a background of the other, which give rise to the 
transitions between the two phases).  We 
then go though the ensemble of states that enter into our averages, 
and probabilistically assign each state to the pinned phase, 
the floating phase, or neither. With this deconvolution, we then 
extrapolate to the $T_c(L)$ which gives equal weight to the two 
phases, and we can then compute the widths $\sigma_{\rm float}$, 
$\sigma_{\rm pin}$, and $\sigma_{\rm tot}$ at this temperature.  Our result 
for $T_c(L)$ is shown in Fig.\,\ref{f2}$a$.  Our results for the 
widths are shown in Fig.\,\ref{f3}.  We find $\sigma_{\rm tot}\sim$ 
constant, and $\sigma_{\rm float}$, $\sigma_{\rm pin}\sim 1/L$, as expected 
for a first order transition \cite{R10}.  We can also use this 
decomposition to compute the entropy jump {\it per vortex} at the transition,
$\Delta s=(\langle E\rangle_{\rm float}-\langle E\rangle_{\rm pin})/(fT_c)$.  Our 
results, shown in Fig.\,\ref{f4}, give a $\Delta s(L)$ that saturates 
to a constant as $L$ increases, again as expected for a 1st order 
transition.
%%
%************ NEW COMES NEXT ***********

\begin{figure}
\epsfxsize=3.2truein
\epsfbox{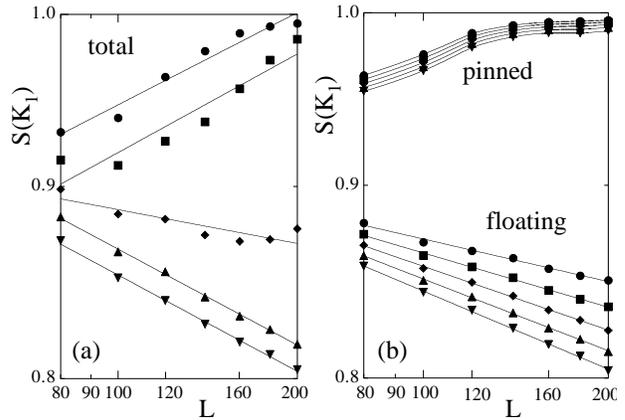}
\vspace{9pt}
\caption{
Log-log plot of $S({\bf K}_1)$ vs. $L$, for several values of
temperature near $T_c$. (a) Shows $S({\bf K}_1)$ averaged over
{\it all} configurations; (b) shows $S({\bf K}_1)$ averaged
separately over only the pinned and only the floating configurations.
From top to bottom the temperatures are, $T=0.00222$, $0.00226$, 
$0.00230$, $0.00234$ and $0.00236$.
}
\label{f5}
\end{figure}

\begin{figure}
\epsfxsize=3.2truein
\epsfbox{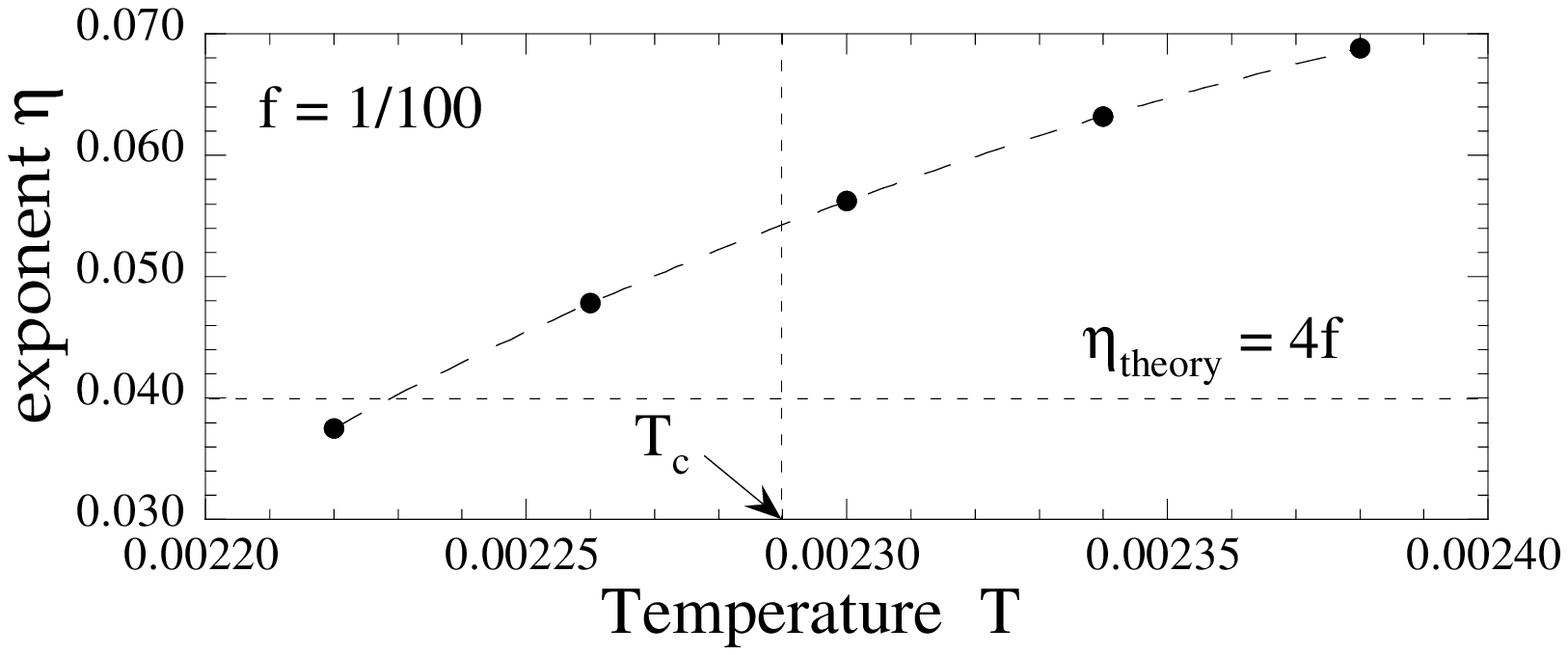}
\vspace{9pt}
\caption{
Translational correlation exponent $\eta(T)$ of the floating
lattice phase, as obtained from the fits in Fig.\,\ref{f5}.
The vortex density is $f=1/100$.
$\eta(T_c)$ remains above the 
theoretical lower limit $4f$ of the Nelson Halperin theory. 
}
\label{f6}
\end{figure}

Next we consider the translational correlations. In the floating solid phase
correlations are algebraic \cite{R2}, $\langle \exp [i{\bf 
K}\cdot ({\bf r}_i-{\bf r}_j)]\sim |{\bf r}_i-{\bf r}_j|^{-\eta_K(T)}$,
with ${\bf K}$ a reciprocal lattice vector, and $\eta_K\equiv \eta |{\bf 
K}|^2/|{\bf K}_1|^2$, with ${\bf K}_1$ the smallest non-zero ${\bf K}$.
Substituting this into the structure function, Eq.(\ref{e2}) results 
in,
%
%****
\begin{equation}
    {S({\bf K})\over L^2} \sim L^{-\eta_K} \sim {\rm e}^{-(\eta\ln L) |{\bf 
K}|^2/|{\bf K}_1|^2}\enspace .
\label{e3}
\end{equation}
%***
%
According to the NH theory \cite{R2}, $\eta$ should take a 
discontinuous jump from zero in the pinned solid to a universal 
finite value at $T_c$ in the floating solid.  For an incompressible 
system such as ours, in which the bulk modulus $\lambda=\infty$, the 
predicted jump is,
%
%***
\begin{equation}
    \eta (T_c^+)=4f\enspace,
    \label{e4}
\end{equation}
%
%****
where $f$ is the vortex density.  From Eq.(\ref{e3}) we see 
that $S({\bf K}_1)/L^2$ vs. $L$ plotted
on a log-log scale should yield a straight line of negative
slope $\eta(T)$.  In Fig.\,\ref{f5}a we show such plots for several
values of $T$ near $T_c$, using $S({\bf K}_1)$ averaged over {\it all}
configurations encountered in the simulation.  The straight lines
are a least squares fit to the assumed algebraic form.  We see that
the fit is not very good, except perhaps at the highest temperatures.
If, however, we average $S({\bf K}_1)$ separately over only
the states in the floating solid phase, and over only the states in the
pinned solid phase, we see the expected behavior as shown in 
Fig.\,\ref{f5}b.  In the pinned phase, the curves
saturate to a finite value as $L$ increases, reflecting the long
range order of this phase.  In the floating phase, the curves give good
straight line fits, showing algebraically decaying correlations.
From these fits, we extract the exponent $\eta$ of the floating phase,
which we plot vs. $T$ in Fig.\,\ref{f6}.  We see that $\eta (T_c^+)$
is close to, but noticeably above, the NH prediction of Eq.(\ref{e4}).
Since the NH mapping onto a vector
Coulomb gas $inverts$ the temperature scale, this is consistent with a 
1st order transition pre-empting the NH defect unbinding transition, 
with Eq.(\ref{e4}) serving as a lower bound on $\eta(T)$. 


\begin{figure}
\epsfxsize=3.2truein
\epsfbox{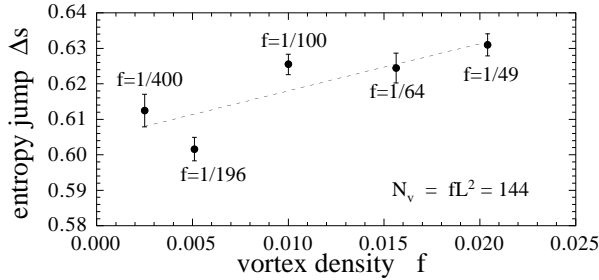}
\vspace{9pt}
\caption{
Entropy jump {\it per vortex}, 
$\Delta s = (\langle E\rangle_{\rm float}-\langle 
E\rangle_{\rm pin})/(fT_c)$,
at the depinning transition vs. vortex density $f$, for systems with a 
fixed number of vortices $N_v=fL^2=144$.  The dashed line is a least
squares fit showing only a weak decrease in $\Delta s$ as $f$ 
decreases.
}
\label{f7}
\end{figure}

The results above are all consistent with a 1st order transition rather than the 
2nd order NH prediction.  However one can question whether this is an 
artifact of the particular density $f=1/100$ that we have used.  
Hattel and Wheatley \cite{R11} in particular have argued that, as $f$ 
decreases in the vortex system, the core energy of the defects that 
lead to depinning increases, and hence even if depinning is 1st order 
at some $f$, it must ultimately become 2nd order as $f$ decreases.  
To check this prediction we have carried out simulations for a variety 
of densities $f=49$, $64$, $100$, $196$, $400$.  Rather than do finite size 
scaling for each case, we study only systems with a fixed total 
number of vortices $N_v$, corresponding to system sizes 
$L=\sqrt{N_v/f}$.  We choose $N_v=144$ because our finite size 
studies of $f=1/100$ indicated this to be sufficiently large to be 
within the asymptotic large $L$ limit.  Analyzing the data from these 
simulations in exactly the same manner as described above, we show in 
Fig.\,\ref{f7} our result for the entropy jump {\it per vortex}
at depinning, $\Delta s$, as a function of density $f$.  
We see that $\Delta s$ is 
only weakly dependent on $f$, extrapolating to a finite 
$\Delta s\simeq 0.60$ as $f\to 0$.  

\begin{figure}
\epsfxsize=3.2truein
\epsfbox{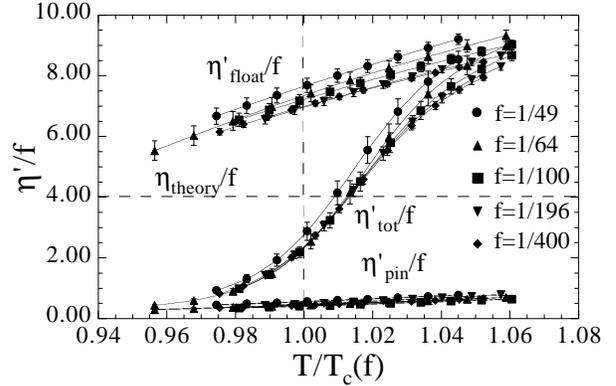}
\vspace{9pt}
\caption{
Estimated exponent $\eta_{\rm tot}^\prime/f$, $\eta_{\rm 
pin}^\prime/f$, and $\eta_{\rm float}^\prime/f$, as determined from 
$S({\bf K})$ averaged over all states, the pinned 
solid states, and the floating solid states, respectively, vs. 
$T/T_c(f)$ 
for different vortex densities $f$.  The data collapse to a common
curve as $f$ decreases. 
}
\label{f8}
\end{figure}

Finally, we consider the translational correlations for the
different $f$.  Since we have simulated only for fixed values of
$fL^2$, we cannot extract $\eta$ by scaling with $L$ as in 
Fig.\,\ref{f5}b.  However we can estimate $\eta$ as follows.
From Eq.(\ref{e3}) we expect a Gaussian envelope for the peaks 
$S({\bf K})$ as a function of $|{\bf K}|$; the width gives a measure
of $\eta$.  This envelope is not purely Gaussian; a correction
exists due to a $|{\bf K}|$ dependence of a prefactor in the
$L^{-\eta_K}$ scaling of Eq.(\ref{e3}).  Making a simple approximate
correction for this effect, as done in Ref.\,\cite{R6}, we plot
our {\it estimate} for the correlation exponent, $\eta^\prime$,
in Fig.\,\ref{f8}.  We show results using $S({\bf K})$ averaged
over all states, only the floating states, and only the pinned states,
plotting the data in scaled units of $\eta^\prime/f$ vs. $T/T_c$.
For $f=1/100$, our estimate $\eta^\prime_{\rm float}$ is slightly larger than
the more correct determination in Fig.\,\ref{f6}.  However the
main point of Fig.\,\ref{f8} is to observe that the estimates
$\eta^\prime$ all collapse to a common curve as the density $f$ 
decreases.  
There is no evidence for a continuing
decrease in the value of $\eta(T_c^+)$ as 
$f$ decreases, as might be the case if the 1st order transition was 
weakening and the NH prediction of Eq.(\ref{e4}) was approached more 
closely.  

To conclude, a finite size scaling analysis for the density $f=1/100$ 
is strongly consistent with the depinning transition being 1st order.
We further find no evidence that this 1st 
order transition changes in any significant 
way if the density $f$ is decreased.  We cannot rule out the 
possibility of different behavior for softer interactions than the 
logarithm considered here.

We would like to thank D. R. Nelson and M. Franz for very helpful 
conversations.  This work was supported by the
Engineering Research Program of the Office of Basic Energy Sciences 
at the Department of Energy, grant 
DE-FG02-89ER14017.

\end{document}